\documentclass[12pt]{iopart}

\begin{document}

\title[Anomalies]{Anomalies, boundaries and the in-in formalism}

\author{Ian G. Moss}
\address{School of Mathematics and Statistics, Newcastle University,
Newcastle upon Tyne, NE1 7RU, UK.}

\begin{abstract}
In the context of quantum field theory, an anomaly exists when a theory has a classical symmetry
which is not a symmetry of the quantum theory. This short exposition aims at introducing a 
new point of view, which is that the proper setting for anomaly calculations is the `in-in', 
or closed-time path formulation of quantum field
theory. There are also some new results for anomalies in the context of boundary value problems,
and a new correction to the $a_5$ heat-kernel coefficient.
\end{abstract}


\section{Anomalies}

In the context of quantum field theory, an anomaly exists when a theory has a classical symmetry
which is not a symmetry of the quantum theory. 
This short exposition aims at introducing a  new point of view, which is that the proper 
setting for anomaly calculations is the `in-in', 
or closed-time path formulation of quantum field
theory. There are also some new results for anomalies in the context of boundary value problems, 
and a new correction to the $a_5$ heat-kernel coefficient. 

As a specific example of an anomaly, consider the breakdown of local gauge invariance in a theory
with a chiral spinor $\psi_L$ in $m$ (even) dimensions, and Dirac operator
\begin{equation}
D_+=\gamma^a(\nabla_a+A_a).
\end{equation}
The gauge field $A_a=A_a^iT_i$, where $T_i$ generates the Lie algebra of our gauge group $G$.

In order to define the quantum theory we have to consider determinants of operators
like $D_+$. This cannot be done directly, because $D_+\psi_L$ and $\psi_L$ have the opposite
chirality. Instead, the traditional approach (see \cite{Nakahara:1990th} for example) 
has been to introduce a fictitious set of antichiral
fields with Dirac operator
\begin{equation}
D_0=\gamma^a\partial_a.
\end{equation}
The determinant of the operator $D_0 D_+$ can be used to define an effective action because it maps
chiral fields to chiral fields. However, the determinant is not always gauge invariant
and can lead to anomalies. More sophisticated approaches to anomalies frame all this in a more 
mathematically elegant way, but the underlying idea remains the same
\cite{AlvarezGaume:1983ig,AlvarezGaume:1984dr,Witten:1985xe}.  

The physical origin of the extra fermion fields and the operator $D_0$ are mysterious. Even the
existence of a trivial connection in $D_0$ may be problematic. The explanation for introducing extra sets 
of fermion fields becomes clear in the `in-in' formalism. This formalism often provides a way 
to define the quantum theory when the more traditional `in-out' route fails. The in-in generating 
function for chiral currents requires two external gauge fields $A$ and $A'$,
\begin{equation}
e^{iW[A,A']}=\langle 0_{in}\mid T^* e^{i\int H_I(A')dt} T e^{-i\int H_I(A)dt}\mid 0_{in}\rangle
\label{inin}
\end{equation}
where $T$ and $T^*$ denote time-ordering and anti-time ordering respectively and the interaction
Hamiltonian $H_I(A)$ is given by
\begin{equation}
H_I=\int \overline\psi_L\,\gamma^aA_a\,\psi_L d^{m-1} x.
\end{equation}
The `in-in' formalism automatically has two sets of fermion fields-one associated with each 
of the time integrals.

Expectation values for the `in-in' formalism are obtained by functional differentiation with respect
to one of the fields and then setting $A'=A$, for example
\begin{equation}
\langle \bar\psi_L \gamma^a T_i\psi_L\rangle=\left.{\delta W\over\delta A^i}\right|_{A'=A}.
\end{equation}
Notice that we set $A'=A$ only at the end, because from (\ref{inin}), it follows that $W[A,A]=0$. 
The generating function $W[A,A']$ also forms part of
an effective action $\Gamma[A,A']$ which generates effective field equations for the expectation
value of the gauge field. Anomalies in the local Lorentz symmetry can be studied in a similar way by
introducing a generating function $W[\omega,\omega']$, depending on two independent spin
connections $\omega$ and $\omega'$. There are many pedagogical accounts of the `in-in' formalism,
also known as the `closed time path' formalism, for example 
\cite{Calzetta:1986cq,Berges:2004yj,Berera:2008ar}.

The following conventions are used. The Lorentzian metric $g_{ab}$ has signature $(-,+\dots +)$.
The Gamma-matrices satisfy $\{\gamma_a,\gamma_b\}=2g_{ab}$ and 
$\gamma_{a\dots b}=\gamma_{[a}\dots\gamma_{b]}$. The Riemann and extrinsic curvature tensors use
Hawking-Ellis conventions and Lie algebra generators $T_i$ are anti-hermitian.

\subsection{Non-abelian anomalies in the `in-in' formalism}

The generating function of the chiral current can be evaluated using standard path integral methods,
with the result that
\begin{equation}
W[A,A']=-i\log\det (D(A')_+{}^{-1}D(A)_+),
\end{equation}
where $D_+$ acts on chiral spinors $S_+$. These operators are defined with an $i\epsilon$
prescription. A new set of operators $D(\tilde A)_-$
acting on chiral spinors $S_-$ can be inserted into the determinant,
\begin{equation}
W[A,A']=-i\log\det (D(A')_+{}^{-1}D(\tilde A)_-{}^{-1}D(\tilde A)_-D(A)_+).
\end{equation}
The point of this manipulation is to get an expression with Dirac operators $D(\tilde A,A)$,
\begin{equation}
W[A,A']=i\log\det D(\tilde A,A')-i\log\det D(\tilde A,A),
\end{equation}
where $D(\tilde A,A)$ acts on the sum $S_+\oplus S_-$,
\begin{equation}
D(\tilde A,A)=\pmatrix{
0&-D_-(\tilde A)\cr D_+(A)&0\cr
}.
\end{equation}
The expectation value of the chiral current is constructed by differentiating with respect to $A$,
and so only the term containing $D(\tilde A,A)$ is relevant. In principle, the expectation values 
should not depend on the dummy field $\tilde A$, but they may depend on $\tilde A$ due to the 
presence of anomalies. 

The determinants can be defined by using the analytic continuation of 
generalised zeta-functions \cite{Dowker77,Hawking77},
\begin{equation}
\zeta(s,D^2)={\rm tr}(D^{-2s}),\label{defzeta}
\end{equation}
where the trace is taken over the Hilbert space of functions with gauge and spinor indices.
These zeta-functions behave best when the operators are elliptic, and so we arrange this by
transforming $D^2$ inside the trace into an elliptic operator. This is commonly associated with the
replacement of the time variable $t$ by $it$ and the corresponding gamma-matrix $\gamma_0$ by
$-i\gamma_0$. How we interpret this step on a time-dependent background is an important
issue which is often overlooked.  The in-in formalism is favoured for time-dependent
backgrounds, but this aspect is not going to be persued further here.

The determinant is defined by analytic continuation in $s$ to $s=0$, 
\begin{equation}
\log\det D^2=-\zeta'(0,D^2).
\end{equation}
We are concerned especially with the way in which the determinant varies under a gauge
transformation of $A$ (keeping $\tilde A$ fixed). If the gauge transformation with parameter
$\alpha$ is denoted by $\delta_\alpha$, then the gauge anomaly $I(\alpha,\tilde A,A)$ is defined to
be
\begin{equation}
I(\alpha,\tilde A,A)=-\frac12\delta_\alpha\zeta'(0,D^2).
\end{equation}
Local Lorentz anomalies $I(\epsilon,\tilde\omega,\omega)$ are defined in a similar way.
Special cases of the gauge anomaly have their own names:
\begin{itemize}
\item The covariant anomaly $I(\alpha,A,A)$;
\item The consistent anomaly $I(\alpha,0,A)$.
\end{itemize}
The full anomaly is similar to an anomaly known as the `VA' anomaly, although the context is rather
different.

For explicit calculations we can replace the definition (\ref{defzeta}) with
\begin{equation}
\zeta(s,D^2)={1\over \Gamma(s)}\int_0^\infty dt\,t^{s-1}{\rm tr}(e^{-(D^2+\epsilon^2)t}).
\end{equation}
The parameter $\epsilon$ takes care of any zero-modes. Zero modes cause the determinant to vanish in
the $\epsilon\to 0$ limit, but the operator expectation values are still well-defined and these are
evaluated in this limit.

The dummy field $\tilde A$ is held fixed whilst taking the gauge variation of $D$, so that
\begin{equation}
\delta_\alpha D=\frac12(1-\tau)\,[D,\alpha],
\end{equation}
where $\tau\equiv \gamma_{l+1}$,
\begin{equation}
\tau=\pmatrix{1&0\cr 0&-1}.
\end{equation}
The anomaly is then
\begin{equation}
I(\alpha,\tilde A,A)=\left.{1\over \Gamma(s)}\int_0^\infty dt\, 
t^{s-1}{\rm tr}(\alpha\tau e^{-(D^2+\epsilon^2) t})\right|_{s=0,\epsilon=0}.\label{iaa}
\end{equation}
where analytic continuation has to be used to define the value at $s=0$.

Applying a second gauge variation to the anomaly (\ref{iaa}) gives the Wess-Zumino consistency
relation
\begin{equation}
\delta_{\alpha_1} I(\alpha_2,\tilde A,A)-\delta_{\alpha_2} I(\alpha_1,\tilde A,A)
=I([\alpha_1,\alpha_2],\tilde A,A).\label{cons}
\end{equation}
The Wess-Zumino consistency condition can be `integrated' to obtain the non-abelian anomaly. Note
that, since $\tilde A$ is inert during all these gauge
transformation, the consistent anomaly
satisfies the Wess-Zumino consistency relation but the covariant anomaly does not. 

The case $I(1,A,A)$ deserves special attention,
\begin{equation}
I(1,A,A)=\left.{1\over \Gamma(s)}\int_0^\infty dt\, 
t^{s-1}{\rm tr}(
e^{-(D_-D_++\epsilon^2) t}-e^{-(D_+D_-+\epsilon^2) t})\right|_{s=0,\epsilon=0}.\label{iaa}
\end{equation}
Since $D_+D_-u=\lambda u$ implies $(D_-D_+)D_-u=\lambda D_-u$, the non-zero spectra of $D_+D_-$ and
$D_-D_+$ are identical and give no net contribution to $I(1,A,A)$. All that remains are the zero modes, 
and these define the twisted Dirac index,
\begin{equation}
I(1,A,A)=n_+-n_-\equiv{\rm index}(D),
\end{equation}
where $n_-$ is the number of zero modes of $D_-$ and $n_+$ is the number of zero modes of $D_+$.

The remarkable feature of the Dirac index is that it can also be expressed in terms of the gauge 
field curvature ${\cal F}$ and the curvature 2-form  ${\cal R}$ of the spacetime 
manifold ${\cal M}$ \cite{Atiyah:1984tf}. 
The result involves two important tensor combinations, the Dirac genus $\hat A(TM)$ and the Chern
character ${\rm ch}({\cal F})$, defined by
\begin{eqnarray}
\hat A(TM)&=&\prod_{j=0}^m{{x_j}/2\pi\over \sinh({x_j}/2\pi)},\\
{\rm ch}({\cal F})&=&{\rm tr}\,\exp(i{\cal F}/2\pi),
\end{eqnarray}
where $x_j$ are the eigenvalues of ${\cal R}$. A term like ${\cal F}^n$ is understood to contain 
both matrix products and exterior products between the
factors. The Dirac index is given by
\begin{equation}
{\rm index}(D)=\int_{\cal M}\,\hat A(TM){\rm ch}({\cal F}),\label{asit}
\end{equation}
where it is meant to be understood that only the $m-$form part of the integrand contributes. 
This index theorem is a special case of
the Atiyah-Singer index theorem, and it can be obtained directly by expanding the operator 
$\exp(-D^2 t)$ for small $t$ as an asymptotic series (see e.g. \cite{Gilkey:1984}).

\subsection{Manifolds with boundary}

Anomalies on odd-dimensional manifolds with boundaries play an important role in $M$-theory
\cite{Horava:1996ma}, and yet
they have been studied much less than their even-dimensional counterparts. 
In even dimensions, the Dirac operator does not always have local boundary conditions 
and so most attention in this area has been on non-local boundary conditions \cite{Gilkey:1984}. 
In odd dimensions, local boundary conditions are possible and
there is a simple index theorem due to Dan Freed \cite{Freed:1996dn}, which is re-derived 
using heat kernel methods below. 

Suppose that the manifold ${\cal M}$ has a spatial boundary $\partial {\cal M}$, meaning that the
normal vector $n$ to the surface is spacelike. Divide the boundary into disjoint components
$\partial {\cal M}_i$ on which a parameter $\epsilon_i$ takes a value $+1$ or $-1$. 
Local boundary conditions can be defined as follows,
\begin{eqnarray}
\psi\in B_{\epsilon}&\hbox{ if }&(1-\epsilon_in^a\gamma_a)\psi=0\hbox{ on }\partial{\cal M}_i,\\
\psi\in B_{\bar\epsilon}&\hbox{ if }&(1+\epsilon_in^a\gamma_a)\psi=0\hbox{ on }\partial{\cal M}_i.
\end{eqnarray}
Exactly as before, the in-in expectation values are generated by
\begin{equation}
W[A,A']=-i\log\det (D(A')_{}^{-1}D(A)),
\end{equation} 
where $D$ now acts on spinors with boundary conditions $B_\epsilon$.
We introduce the dummy field
$\tilde A$, and focus on
\begin{equation}
D(\tilde A,A)=\pmatrix{
0&-D(\tilde A)\cr D(A)&0\cr}.
\end{equation}
acting on $B=B_{\epsilon}\oplus B_{\bar\epsilon}$. The gauge transformations lead to an anomaly in
$m=2l+1$ dimensions given by
\begin{equation}
I(\alpha,\tilde A,A)=\left.{1\over \Gamma(s)}\int_0^\infty dt\, 
t^{s-1}{\rm tr}(\alpha \tau e^{-(D^2+\epsilon^2) t})\right|_{s=0,\epsilon=0},
\end{equation}
with boundary conditions $B$. Similarly, the index of the Dirac operator with boundary conditions
$B_\epsilon$ is
\begin{equation}
\hbox{index}(D,B_\epsilon)=\left.{1\over \Gamma(s)}\int_0^\infty dt\, 
t^{s-1}{\rm tr}(\tau e^{-(D^2+\epsilon^2) t})\right|_{s=0,\epsilon=0}.\label{indodd}
\end{equation}
The index can be used to generate the anomaly, and so we let $\tilde A=A$ from this point on.

As a first example, consider the product manifold ${\cal M}=X\times Y$, with the interval $Y=[0,1]$
and the Dirac operator $D=D_X\otimes1\oplus1\otimes D_Y$. Choose boundary conditions with
$\epsilon_1=1$ and $\epsilon_0=-1$ at the two ends of the interval. The two boundaries are
$\partial {\cal M}_1\simeq X$ and $\partial {\cal M}_0\simeq -X$, where the minus sign indicates a
change of orientation of the spin basis.

Suppose that $D\psi=0$ and $\psi\in B_\epsilon$. Note that the Hermitian product 
$0=\langle D\psi,D\psi\rangle=\langle D_Y\psi,D_Y\psi\rangle+\langle D_X\psi,D_X\psi\rangle$. Both
terms are positive, and so $\psi$ is also a zero mode of $D_X$. The boundary conditions
$B_\epsilon$ imply that $\psi$ is chiral on $\partial{\cal M}_1$ and antichiral on 
$\partial{\cal M}_0$. The converse of this argument also applies,
so we conclude that the zero modes are in 1-1 correspondence. Also
$\hbox{index}(D_{\partial{\cal M}_1})=-\hbox{index}(D_{\partial{\cal M}_0})$, hence
\begin{equation}
\hbox{index}(D,B_\epsilon)=\frac12\sum_i\epsilon_i\,\hbox{index}(D_{\partial{\cal M}_i})\label{dpf}
\end{equation}
This is the simple version of the index theorem (i.e. with vanishing extrinsic curvature) 
first used by Witten to demonstrate anomaly cancellation in heterotic $M$-theory \cite{Horava:1996ma}.

Before proceding to the general case, it is useful to introduce some new notation. Let  ${\cal
D}=\nabla+A$ and introduce new
gamma-matrices for
the enlarged spinor space, 
\begin{equation}
\hat\gamma_a=\pmatrix{\gamma_a&0\cr 0&-\gamma_a},\quad
I=\pmatrix{0&1\cr 1&0},\quad\chi_i=\epsilon_i\pmatrix{n^a\gamma_a&0\cr 0&-n^a\gamma_a}.
\end{equation}
In this notation the operator $D=I\hat\gamma^a{\cal D}_a$.

The index formula (\ref{indodd}) requires boundary conditions on the normal derivatives. These
boundary conditions can be obtained by starting from
\begin{equation}
P_\epsilon\psi=P_{\epsilon} D\psi=0\hbox{ on }\partial {\cal M}.
\end{equation}
where
\begin{equation}
P_\epsilon=\frac12(1- \chi).
\end{equation}
Note that
\begin{equation}
\left\{\chi,\hat\gamma^b{\cal D}_b\right\}=2n^a{\cal D}_a+k\epsilon,
\end{equation}
where $k$ is the trace of the extrinsic curvature. It follows that
\begin{equation}
P_{\epsilon}I\hat\gamma^a{\cal D}_a\psi=I(n^a{\cal D}_a+\frac12k\epsilon)P_{\bar\epsilon}\psi.
\end{equation}
Therefore the boundary conditions for $D^2$ are
\begin{equation}
P_\epsilon\psi=(n^a{\cal D}_a+\frac12k\epsilon)P_{\bar\epsilon}\psi=0\hbox{ on }\partial {\cal M}.
\end{equation}
Boundary conditions like these are known as mixed boundary conditions 
\cite{Luckock:1989jr,Moss:1994jj,Dowker:1995sw}.

The local expression for the Dirac index can be obtained from a heat kernel expansion on the
manifold with boundary. In $m=2l+1$ dimensions,  this takes the form \cite{Gilkey:1984}
\begin{equation}
{\rm tr}(f e^{-D^2 t})\sim\sum_{n=0}^\infty a_n(f,D^2,B)t^{(n-m)/2},
\end{equation}
where the coefficients $a_n(f,D^2,B)$ are integrals of local invariants. There
are no interior terms for $n$ odd, and only the boundary contributes.
The index obtained from (\ref{indodd}) is then
\begin{equation}
{\rm index}(D,B_\epsilon)=a_{m}(\tau,D^2,B).
\end{equation}
The heat kernel expansion has been studied for operators of the form $-{\cal D}^2-E$ and mixed
boundary conditions $P_\epsilon\psi=(n^a{\cal D}_a-S)P_{\bar\epsilon}\psi=0$.  The invariants in
the boundary coefficients are combinations of $\chi$, $S$, $E$, the curvature 
$\Omega_{ab}=[{\cal D}_a,{\cal D}_b]$ and tangential covariant derivatives 
\cite{Luckock:1989jr,Moss:1989mz,Branson:1999jz}.

Most of the invariants do not contribute to the index of the Dirac operator due to a special
property of traces over the spinor indices. Suppose that $a_1\dots a_{2l}$ are tangential indices
ordered to be consistent with the orientation of the normal vector, then
\begin{equation}
\hbox{tr}(\tau
\hat\gamma_n\hat\gamma_{a_1}\dots\hat\gamma_{a_{2l}})=(-i)^l 2^{l+1}
\varepsilon_{a_1\dots a_{2l}}.\label{trg}
\end{equation}
but {\it the traces of all other combinations of gamma-matrices vanish}.

Tensors with gamma-matrices in the heat kernel coefficients of $D^2$ are
\begin{eqnarray}
\Omega_{ab}&=&\frac14R_{abcd}\hat\gamma^{cd}+F_{ab}\label{omega}\\
E&=&-\frac12F_{ab}\hat\gamma^{ab}-\frac14R\\
\chi_{|a}\chi_{|b}&=&k_{ac}k_{bd}\hat\gamma^c\hat\gamma^d\label{chi}
\end{eqnarray}
We require at least $l$ of these tensors or their derivatives for a non-vanishing contribution to
$a_m$, but $l$ such terms already have the maximal (conformal) dimension allowed for $a_m$.
Tangential derivatives increase the dimension and therefore terms with more derivatives are not
allowed. Terms with $S$ also increase the dimension and these are not allowed. 

After all the allowed contributions have been combined, the Gauss-Codacci relations can be used to
replace $R_{abcd}$ with the surface curvature components $r_{abcd}$, the extrinsic curvature $k$ and
its tangential derivative. Then,
\begin{equation}
a_m=\sum_n\int_{\partial{\cal M}}\epsilon\, P_n(r,F,k),
\end{equation}
where $P_n$ is a polynomial of degree $n$ in $k$. Comparison of the
direct product formula (\ref{dpf}) and the index theorem (\ref{asit}) gives the result when $k=0$,
$2P_0=\hat A(TM){\rm ch}(F)$. For the other terms, we
note that it is possible to deform the metric on the manifold to obtain a new metric with the same
boundary geometry but with extrinsic curvature $\lambda k_{ab}$, where $\lambda$ is a real number.
The index with the new metric is
\begin{equation}
\hbox{index}(D,B_{\epsilon})=
\sum_n\lambda^n\int_{\partial{\cal M}}\epsilon\, P_n(r,F,k).
\end{equation}
The left-hand side of this expression is still an integer, and so all of the terms on the right
apart from the first one must vanish. The integral of $P_0$ reproduces the Dirac index on the
boundary, hence  
\begin{equation}
\hbox{index}(D,B_{\epsilon})=\frac12\sum_i\epsilon_i\,\hbox{index}(D_{\partial{\cal M}_i})
\label{findex}
\end{equation}
A similar argument can be appled to Rareta-Schwinger fields.

The gauge anomaly $I(\alpha,\tilde A, A)$ and the local lorentz anomaly 
$I(\epsilon,\tilde\omega,\omega)$ can be obtained from the index using the methods of a later
section. It follows, in particular, that these anomalies depend only on intrinsic properties of the
boundary. An application of these results to heterotic $M$-theory is given in Ref. \cite{Moss:2005zw}.

\subsection{Boundary anomalies in five dimensions}

All of the relevant heat kernel coefficients are known in five dimensions, enabling us
to illustrate the general procedure described above in a particular example. It turns out
that the calculation fails due to an apparent error in one of the heat kernel coeffiiecents derived
in Ref. \cite{Branson:1999jz}, and we will use the anomaly calculation to correct this coefficient.

Eq. (\ref{trg}) says that only the terms in $a_5$ which contain one normal and four 
tangential gammas can contribute. These are
\begin{eqnarray}
a_5(f,D^2,B)&=&{1\over 16\pi^2}{1\over 5760}\int_{\partial{\cal M}}{\rm tr}f\left(
720\chi E^2+w_6\chi\Omega_{ab}\Omega_{ab}\right.\nonumber\\
&&\left.+w_9\chi\Omega_{an}\Omega_{an}+w_{14}\chi\chi_{|a}\chi_{|b}\Omega_{ab}
+w_{20}\chi\chi_{|a}\chi_{|a}E\right).
\end{eqnarray}
The coefficients $w_i$ are labelled so as to agree with Ref. \cite{Branson:1999jz},
who give $w_6=120$, $w_9=180$ and $w_{14}=90$.
For the Dirac operator with (\ref{omega}-\ref{chi}),
\begin{eqnarray}
{\rm index}(D,B_\epsilon)&=&{1\over 16\pi^2}{1\over 5760}\int_{\partial{\cal M}}\left(
-1440\epsilon_{abcd}{\rm tr}\,F_{ab}F_{cd}-\frac12 w_6\epsilon_{abcd}R_{abef}R_{cdef}\right.
\nonumber\\
&&\left.+\frac12w_9\epsilon_{abcd}R_{aben}R_{cden}
+2 w_{14}\epsilon_{abcd}k_{ae}k_{bf}R_{cdef}\right).
\end{eqnarray}
This can be expressed in terms of surface curvature $r_{abcd}$ by using Gauss-Codazzi relations,
\begin{eqnarray}
R_{abcd}&=&r_{abcd}-k_{ac}k_{bd}+k_{ad}k_{bc},\\
R_{abcn}&=&k_{be|a}-k_{ae|b}.
\end{eqnarray}
There is a useful identity,
 \begin{equation}
\epsilon_{abcd}k_{ae}k_{bf}r_{cdef}=-2\epsilon_{abcd}k_{ae|bd}k_{ce}.
\end{equation}
The index becomes
\begin{eqnarray}
{\rm index}(D,B_\epsilon)&=&{1\over 16\pi^2}\int_{\partial{\cal M}}\left(
-{\rm tr}({\cal F}^2)+{w_6\over 2880}{\rm tr}({\cal R}^2)\right.\nonumber\\
&&\left.+\frac1{2880}(2w_6-2w_{14}-w_9)\epsilon_{abcd}k_{ae|b}k_{ce|d}\right).
\end{eqnarray}
Comparison with the index theorem (\ref{findex}) and the index formula (\ref{asit})
gives $w_6=120$, in agreement with Ref. \cite{Branson:1999jz}.  However, the
remaining coefficients leave an extrinsic curvature term which is
inconsistent with the index theorem.  

The disagreement can be rectified by correcting Lemma 7.1 in Ref. \cite{Branson:1999jz}.
The correct version of the table which appears in the proof of Lemma 7.1 is given below:

\begin{table}[ht!]
\begin{center}
\begin{tabular}{lcc}
\hline
Invariant&Coefficient of $f_m^2f_{a|a}$&Coefficient in $a_5$\\
\hline\hline     
$\chi\chi_a\chi_a\Omega_{ab}$&$-8(m-2)$&$w_{14}$\\
$\chi E^2$&$-2(m-1)$&720\\
$\chi_aE_a\ (+t.d)$&$-2(m+1)$&-180\\
$\chi\chi_a\chi_aE$&$-4(m-1)$&-90\\
$\chi\Omega_{ab}\Omega_{ab}$&$8(m-2)$&120\\
\hline                        
\end{tabular}
\label{table}
\end{center}
\end{table}

The coefficient of $f_m^2f_{a|a}$ vanishes, and we deduce that $w_{14}=30$ (not 90). 
Taking the other coefficients as being correct, then $2w_6-2w_{14}-w_9=0$ and 
the extrinsic curvature terms do not contribute to the index.

\subsection{Non-abelian anomalies}

A standard approach developed by Wess and Zumino allows us to obtain the local anomaly by 
dimensional reduction of the index formula {(see e.g. \cite{Nakahara:1990th})}. 
Suppose that the boundary has $2l$ dimensions. First, write the index for $2l+2$ 
dimensions as the integral of a $d+2$
form $I_{2l+2}$ (as in Eq. (\ref{asit})). Next, find the transgression $TI_{2l+2}$, which
is a $2l+1$-form with the property that $d\,TI_{2l+2}=I_{2l+2}$. Finally, solve
the equation $dQ=\delta_\alpha TI_{2l+2}$. The anomaly is given by
\begin{equation}
I(\alpha,0,A)=\frac12\sum_i\epsilon_i\int_{\partial M_i} Q_i(\alpha,0,A)\label{defg}
\end{equation}
for a suitable normalisation of the $Q_i$. 

It is simple to check whether the consistency relation (\ref{cons}) is satisfied. If the boundary
components are suitably chosen with $Q_i=\epsilon_iQ$, we have
\begin{equation}
I(\alpha,0,A)=\frac12\sum_i\epsilon_i\int_{{\cal \partial M}_i}Q_i(\alpha,0,A)=
\int_{\cal M}dQ(\alpha,0,A)=\delta_\alpha\int_{\cal M}\,TI_{d+2}.
\end{equation}
The consistency condition then follows simply from
\begin{equation}
[\delta_{\alpha_1},\delta_{\alpha_2}]=\delta_{[\alpha_1,\alpha_2]}.
\end{equation}
The anomaly depends only on the geometry of the boundary. This raises an unresolved issue
in Heterortic $M$-theory where a Green-Squartz mechanism \cite{green} is used to 
cancel the anomalies, but the cancellation is only modulo extrinsic curvature terms. 

\subsection{Solving the Wess-Zumino consistency condition}

We have seen how the non-abelian anomaly $I(\alpha,\tilde A,A)$ arises in the context of the `in-in'
formalism. In this section we shall calculate $I(\alpha,\tilde A,A)$ in flat spacetime by solving
the consistency condition. The steps are based on the standard approach  
(see e.g. \cite{Nakahara:1990th}).  There is a deep topological magic underlying the 
process, but we need only follow some simple algebraic steps. An advantage of having $\tilde A\ne0$ 
is that the index $I(1,A,A)$ can be used to fix the normalisation of the anomaly.

The solution to the consistency condition starts with the definition of the transgression of the
Chern characters,
\begin{equation}
Tch_{l+1}(\tilde A,A)=
{1\over l!}\left({i\over 2\pi}\right)^{l+1}\int_0^1dt\,{\rm str}\left(A-\tilde A,F_t,\dots
F_t\right),\label{deftch}
\end{equation}
where `str' denotes a symmetrised local trace of the $j+1$ factors and 
\begin{equation}
A_t=\tilde A+t(A-\tilde A),\qquad F_t=dA_t+A_t^2.
\end{equation}
Note that the dimension is effectively $2l+2$ at this stage. The point of introducing the
transgression is that it can be shown to satisfy
\begin{equation}
d \,Tch_{l+1}(\tilde A,A)=ch_{l+1}(F)-ch_{l+1}(\tilde F),
\end{equation}
where $ch_j(F)$ is the $j$'th Chern-character,
\begin{equation}
ch_{l+1}(F)={1\over (l+1)!}\left({i\over 2\pi}\right)^{l+1}{\rm tr}\left(F^{(l+1)}\right).
\end{equation}
The next step is to use the solution to the consistency conditions in $m=2l$ dimensions  given by
\begin{equation}
I(\alpha,\tilde A,A)=\int_{\cal M}Q(\alpha,\tilde A,A),
\end{equation}
where $Q(\alpha,\tilde A,A)$ is a solution to
\begin{equation}
dQ(\alpha,\tilde A,A)=\pi \,\delta_\alpha Tch_{l+1}(\tilde A,A).\label{defq}
\end{equation}
The gauge variation applies to the $A$ but not the $\tilde A$. 

By adapting the work of Wess and Zumino, it is possible to show that 
$-2\pi i\,Tch_{l+1}(\tilde A,A+\alpha)$,
when expanded to linear order in $\alpha$, automatically solves (\ref{defq}). After doing this
replacement in (\ref{deftch}), and picking a normalisation factor, we obtain
\begin{equation}
Q(\alpha,\tilde A,A)=
{1\over l!}\left({i\over 2\pi}\right)^l\int_0^1dt\, {\rm str}\left(
\alpha F_t^l+l \, t(t-1)\dot A_t[\alpha,\dot A_t] F_t^{l-1}\right)
\end{equation}
where $\dot A_t=A-\tilde A$.
Some points to note about this expression are:
\begin{enumerate}
\item by changing variable from $t$ to $1-t$, we see that $Q(\alpha,\tilde A,A)$ is symmetric under
the interchange of $\tilde A$ and $A$;
\item when $\tilde A=A$, the result reproduces the covariant anomaly
\begin{equation}
Q(\alpha,A,A)=
{1\over l!}\left({i\over 2\pi}\right)^l\, {\rm str}\left(\alpha F^l\right);
\end{equation}
\item when $\tilde A=A$ and $\alpha=1$, the result correctly produces the index of the Dirac
operator in $m=2l$ dimensions, confirming the choice of normalisation;
\item when $\tilde A=0$, some further manipulation of the expression gives a standard form for the
consistent anomaly,
\begin{equation}
Q(\alpha,0,A)=
{1\over (l-1)!}\left({i\over 2\pi}\right)^l\int_0^1dt\,(1-t) {\rm str}
\left( \alpha d(\dot A_t F_t^{l-1})\right).
\end{equation}
\end{enumerate}
In curved spacetime, the Dirac index is constructed by combining the Chern character with the 
$\hat A$ genus of the metric
connection. The non-abelian
anomaly is quite complicated, and it is often easiest to repeat the steps above on a case by case
basis.

\ack

This paper was prepared in honour of Stuart Dowker's 70'th birthday. It is a pleasure
to know him and an honour to have worked with him over the years.  

The author is supported by the STFC Consolidated Grant ST/J000426/1.

\bibliographystyle{unsrt}
\section*{References}
\bibliography{anomalies.bib}
\end{document}